\documentclass{article}
\usepackage[dvips]{graphicx}

\def\beq{\begin{equation}}            \def\eeq{\end{equation}}
\def\bear{\begin{eqnarray}}           \def\eear{\end{eqnarray}}
\def\la{\langle}    \def\ra{\rangle}  
    \def\ci{\cite}    \def\lb{\label}
\def\td{\tilde}

\def\Dlt{\Delta}     
\def\vphi{\varphi}
\begin{document}

\title{\hfill{\normalsize quant-ph/0209032$\qquad$}\\[3mm]
Comment on "{\it On the uncertainty relations and squeezed states for
the quantum mechanics on a circle}"}
\author{D.A. Trifonov\\
Institute for nuclear research,\\
72 Tzarigradsko chaussee, 1784 Sofia, Bulgaria}
\maketitle
\date{}

\begin{abstract}
It is shown by examples that the position uncertainty on a circle,
proposed recently by Kowalski and Rembieli\'nski [J. Phys. A 35 (2002)
1405] is not consistent with the state localization.  We argue that the
relevant uncertainties and uncertainty relations (UR's) on a circle are
that based on the Gram-Robertson matrix. Several of these generalized UR's
are displayed and related criterions for squeezed states are discussed.
\end{abstract}
\vspace{5mm}

{\bf\large 1.}
In the recent paper \ci{KR} the problem of relevant uncertainty
relation (UR) for the angular momentum and the angle variables of a
particle on a circle was discussed and a new UR was proposed. Noting a
contradiction in the previously obtained UR \ci{GO} the authors defined  new
quantities $\td\Dlt^2(\hat{\vphi})$ and $\td\Dlt^2(\hat{J})$ ($\hat{\vphi} = \vphi$,
$\hat{J} = -id/d\vphi$) as measures for the
uncertainty of the angle $\vphi$ and the angular momentum $J$
and suggest the inequality 
\beq\lb{KR UR}  %eq. (1)
\td\Dlt^2(\hat{\vphi}) + \td\Dlt^2(\hat{J}) \geq 1.
\eeq
The quantities $\td\Dlt^2(\hat{\vphi})$, $\td\Dlt^2(\hat{J})$ are defined
as \ci{KR} (note a change in notation: $\Dlt^2\rightarrow\td\Dlt^2$)
\beq\lb{KR Dlt}  %eq. (2)
\td\Dlt^2(\hat{\vphi}) = -\frac{1}{4}\ln\left|\la U^2\ra\right|^2,\qquad
\td\Dlt^2(\hat{J}) =
\frac{1}{4}\ln\left(\la e^{-2\hat{J}}\ra\la e^{2\hat{J}}\ra\right),
\eeq
where $U=\exp(i\hat{\vphi})$.  The authors of \ci{KR} find that for the
eigenstates $|z\ra$ of the operator $Z=\exp(-\hat{J}+1/2)U$, $Z|z\ra =
z|z\ra$ (the "genuine coherent states (CS's) for a quantum particle on
a circle" \ci{KRP}) both quantities (\ref{KR Dlt}) equal $1/2$,
% \td\Dlt^2(\hat{\vphi})_z = 1/2 = \td\Dlt^2(\hat{J})_z
and suggest that $\td\Dlt^2(\hat{\vphi})$ and
$\td\Dlt^2(\hat{J})$ obey the inequality (\ref{KR UR}) in any state.
Henceforth, the quantities $\td\Dlt^2(\hat{\vphi})$ and $\td\Dlt^2(\hat{J})$
should be referred to as Kowalski-Rembieli\'nski uncertainties ((K-R)
uncertainties), the UR (\ref{KR UR}) -- as K-R UR and $|z\ra$ -- as
Kowalski-Rembieli\'nski-Papaloucas CS's (K-R-P CS's).
\vspace{2mm}

{\bf\large 2.}
Next we shall demonstrate that the K-R uncertainty
$\td\Dlt^2(\hat{\vphi})$ is not consistent with the state localization on
a circle.  For this purpose we compare the $\vphi$--probability
distributions $p_\psi(\vphi)$ (defined as $p_\psi(\vphi) = |\psi(\vphi)|^2
= |\la\vphi|\psi\ra|^2$) in K-R-P CS's with $\vphi$--distributions in
certain states with squeezed $\td\Dlt^2(\hat{\vphi})$.  The quantity
$\td\Dlt^2(\hat{\vphi})$ is called \ci{KR} squeezed if it is less that
than $1/2$. The authors of \ci{KR} constructed a family of such squeezed
states $|z\ra_s$ as eigenstates of the operator $Z(s) =
\exp(-s\hat{J}+s/2)U=\exp(i\vphi-s\hat J)$, where $s$ is positive
parameter.  Here we shall consider $\td\Dlt^2(\hat{\vphi})$--squeezed states
of the form of eigenstates $|z,a\ra$ of the squared operator $Z^2$. These
are defined as macroscopic superpositions of $|z\ra$ and $|-z\ra$
(Schr\"odinger cat states on a circle),
\beq\lb{|z,a>}   %eq. (4)
|z,\,a\ra = N(z,\,a)\,(|z\ra + \,a\,|\!-\!z\ra),
\eeq
where $\,a$ is complex parameter, and the normalization constant
$N(z,\,a)$ takes the form
% (in fact $N$ does not depend on the phase of $z$)
\beq\lb{N}  %eq. (5)
 N(z,\,a) = \left[1+|\,a|^2 + 2\la z|\!-\!z\ra{\rm
Re}\,a\right]^{-1/2}.
\eeq
The scalar product of two CS's is \ci{KR} $\la z|\eta\ra =
\theta_3\left((i/2\pi)\ln(z^*\eta),i/\pi\right)$, where $\theta_3(x,y)$ is
the Jacobi theta-function. The states $|z,a\!=\!\pm1\ra\equiv |z;\pm\ra$
should be called even/odd CS on a circle.

On the states $|z,a\ra$ the quantities $\la U^2\ra$, $\la \exp(2J)\ra$,
$\la\exp(-2J)\ra$  in (\ref{KR Dlt}) take the form
\bear     %eq. (6) (7)
\la\,a,z| U^2|z,\,a\ra \!&=&\!
N^2(z,\,a)(\la z|U^2|z\ra + |\,a|^2\la\!-\!z|U^2|\!-\!z\ra +
\,a\la z|U^2|\!-\!z\ra + \,a^*\la\!-\!z|U^2|z\ra), \lb{KR Dlt 1a}\\
\la\,a,z| e^{\pm2J}|z,\,a\ra \!&=&\!
N^2(z,\,a)\left(\la z|e^{\pm2J}|z\ra +
|\,a|^2\la\!-\!z|e^{\pm2J}|\!-\!z\ra + 2{\rm Re}
\left(\,a\la z|e^{\pm2J}|\!-\!z\ra\right)\right), \lb{KR Dlt 1b}
\eear
where  $\la z|U^2|z\ra$, $\la z|e^{2J}|z\ra$ and $\la z|e^{-2J}|z\ra$  are
given by $z/ez^*$, $e/|z|^2$ and $e|z|^2$ respectively \ci{KR}.
Substituting (\ref{KR Dlt 1a}) and (\ref{KR Dlt 1b}) in (\ref{KR Dlt}) we
obtain explicit formulas for K--R uncertainties in $|z,a\ra$.

From formulas (\ref{KR Dlt}), (\ref{KR Dlt 1a}) and (\ref{KR Dlt 1b})  we
find that $\td\Dlt^2(\hat{\vphi})$--squeezing occurs in many superpositions
$|z,\,a\ra$, in particular in $|z;\pm\ra$ (see figure 1).\, In the odd
state $|1;-\ra$, corresponding to the solid line minimum in figure 1, we
find $\td\Dlt^2(\hat{\vphi}) \approx 0.33$, which is considerably less than
the value $1/2$ of $\td\Dlt^2(\hat{\vphi})$ in CS's $|z\ra$.  One should
expect that the $\vphi$-distribution, corresponding to wave functions with
squeezed "position uncertainty" $\td\Dlt^2(\hat{\vphi})$ are better
localized on the circle than the non-squeezed CS. Unfortunately it is not
the case with $\td\Dlt^2(\hat{\vphi})$--squeezed states from the family
$\{|z,\,a\ra\}$.  This inconsistency is demonstrated in figure 2 on the
example of cat state $|1;-\ra$.  As one can see from figure 2 the
$\td\Dlt^2(\hat{\vphi})$--squeezed state $|1;-\ra$ is much worse localized
than the non-squeezed CS $|z\ra$ ($p(\vphi)$-distributions of $|z\ra$ with
different $z$ approximately coincide up to a translation). Therefore the
quantity $\td\Dlt^2(\hat{\vphi})$ is not a proper measure of the position
uncertainty, and the inequality (\ref{KR UR}) could hardly be
qualified as a relevant uncertainty relation on a circle.

Let us note that $|z,a\ra$ saturate the inequality (\ref{KR UR}) with
unequal $\td\Dlt^2(\hat{\vphi})$ and $\td\Dlt^2(\hat J)$, the case of $z=1$ 
and real $a$ being demonstrated in figure 1. However the whole range of
validity  of (\ref{KR UR}) is not yet clarified. Nevertheless it might be
interesting to note that in the variety of states on the real line a
similar inequality hold, i.e.  $\td\Dlt^2(\hat x)+ \td\Dlt^2(\hat p) \geq
1$, where $\hat x$, $\hat p$ are position and momentum operators.
\vspace{2mm}

{\bf\large 3.}
The above remarks naturally raise again the questions about the position
and angular momentum uncertainties and the relevant {\it uncertainty
relations (UR's) on a circle}. In my opinion most suitable UR's for $n$
observables $X_i$ and one state $|\psi\ra$ on a circle are those based on
the Gram--Robertson matrix $G=\{G_{ij}\}$ of the form \ci{T} ($i,j =
1,\ldots,n$;\, $n=1,2,\ldots$)
\beq\lb{G}    %eq. (8)
G_{ij}(\psi) = \la(X_i-\la X_i\ra)\psi|(X_j-\la X_j\ra)\psi\ra.
\eeq
More informative notations $G(\vec{X};\psi)$ and $G_{X_iX_j}(\psi)$
($\vec{X}=X_1,X_2$) for this matrix and its elements should also be used.
The generalized covariances $\,_g\Dlt X_iX_j(\psi)$ of $X_i$ and $X_j$ in
$|\psi\ra$ are defined \ci{T}  as symmetric
part $S_{ij}$ of $G_{ij}$ (for the case of $n=2$
see also \ci{Ch,D})
\beq\lb{Cov2}   %eq. (9)
_g\Dlt X_iX_j(\psi) := S_{X_iX_j}(\psi) = {\rm Re}\la(X_i-\la
X_i\ra)\psi|(X_i-\la X_i\ra) \psi\ra.
\eeq
The diagonal elements $S_{ii}$ are defined as generalized variances
$(_g\Dlt X_i)^2$ of $X_i$).

Since $G$ is non-negative all the characteristic coefficients of its symmetric
part $S= (G+G^T)/2$ are not less than the corresponding characteristic
coefficients of its antisymmetric part $A = (G-G^T)/2i$.
These inequalities are called generalized {\it
characteristic UR's} \ci{T}. The senior characteristic UR reads
\beq\lb{det UR}  %eq. (10)
\det S(\vec{X};\psi) \geq \det A(\vec{X};\psi)\, .
\eeq
In the simplest case of $n=2$ this UR is displayed as $S_{11}S_{22} -
S_{12}^2 \geq A_{12}^2$. It can also be written in the shorter form
$\det G \geq 0$, and displayed in terms of the generalized
covariances as
\beq\lb{gSUR}   %eq. (11)
(_g\Dlt X_1)^2(_g\Dlt X_2)^2 \,\geq\, (_g\Dlt X_1X_2)^2 + \left({\rm Im}
\la(X_1-\la X_1\ra)\psi|(X_2-\la X_2\ra) \psi\ra\right)^2.
\eeq
The sum of the two terms in the right-hand-side of (\ref{gSUR}) is just
the squared absolute value of $G_{12}$, i.e. we have $_g\Dlt X_1\,_g\Dlt
X_2 \geq |G_{12}(\psi)|$.

When the actions of $X_iX_j$ on $|\psi\ra$ are correctly defined (normal
cases) the above Gram matrix coincides \ci{T} with the Robertson one
\cite{R}: its antisymmetric part $A_{ij}$ reduces, up to a factor, to the
mean commutator, $A_{ij} = -(i/2)\la[X_i,X_j]\ra$, and its symmetric part
takes  the familiar form of the standard uncertainty matrix
$\sigma(\vec{X};\psi)$. (The element $\sigma_{ij}=\la
X_iX_j+X_jX_i\ra/2-\la X_i\ra\la X_j\ra \equiv \Dlt X_iX_j$ is standard
covariance of $X_i$ and $X_j$, and $\sigma_{ii}=\Dlt X_iX_i\equiv (\Dlt
X_i)^2$ is the variance of $X_i$. $(\Dlt X)^2$ should not be
confused with the K-R quantity $\td\Dlt^2(X)$).  Under these conditions the
inequality (\ref{det UR}) takes the form of Robertson UR for $n$ observables
\ci{R,T,T2}, and (\ref{gSUR}) coincides with the Schr\"odinger (or
Schr\"odinger--Robertson) UR \ci{S} (for a review on this UR and its
minimization states see e.g. \ci{T2}).

The generalized form of the less precise Heisenberg UR reads  $\,_g\Dlt
X_1\,_g\Dlt X_2 \geq |{\rm Im}G_{12}|$, and it again follows from the more
precise one (\ref{gSUR}). About similar generalization see also \ci{Ch,D}.
(Please note that in some papers, e.g. \ci{A}, no distinction is made
between Schr\"odinger and Heisenberg UR's, both being named after
Heisenberg).

Thus in the special cases when $X_iX_j|\psi\ra$ are not properly
defined one should resort to generalized Schr\"odinger UR (\ref{gSUR})
(for two observables), and to (\ref{det UR}) (for several observables).
The position and the angular momentum observables of a particle on a circle
represent such a special case, since $\hat{\vphi}\la\vphi|\psi\ra =
\vphi\la\vphi|\psi\ra$ is not $2\pi$-periodic and $\hat{J}$ is not
Hermitian on such functions. Another special case of interest is particle
motion on the sphere.

Figure 3 illustrates the generalized UR (\ref{gSUR}) in the case of
$X_1=\hat{J}$ and $X_2=\hat{\vphi}$  and states $|z,a\ra$ (particle on a
circle), where $\det G(z,a) = \det G(\vec{X};z,a)$ are plotted as
functions of real $a$ for $z=0.4$ (solid line) and $z=1$ (dashed line).
In these states the generalized covariance $\,_g\Dlt\vphi J = {\rm
Re}G_{J\vphi}$ vanishes, also $\,_g\Dlt\vphi=\Dlt\vphi$, $\,_g\Dlt J=\Dlt
J$, so that here we have $\det G = (\Dlt\vphi)^2(\Dlt J)^2 - ({\rm
Im}G_{J\vphi})^2 \geq 0$.  The minimal value of $\det G$ on figure 3 is
different from zero (it is about $0.00017$).

Unlike $\td\Dlt^2(\hat{J})$ and $\td\Dlt^2(\hat{\vphi})$, the
variances $(\Dlt J)^2$ and $(\Dlt \vphi)^2$ are in good correspondence
with the angular momentum and position localization on a circle. For
example, $\vphi$-distributions for CS's $|z\ra$ with $z=0.4,\,1$
practically are the same (see figure 2), and the variances $(\Dlt\vphi)^2$
in these CS are almost equal: in $|z\!=\!0.4\ra$\, $(\Dlt\vphi)^2 =
0.50055$, and in $|z\!=\!1\ra$\, $(\Dlt\vphi)^2 = 0.50064$. In the worse
localized cat state $|1;-\ra$ (see figure 2) the variance
$(\Dlt\vphi)^2$ takes the larger value of $3.813$.

We have to warn that one has to be careful about the
correspondence between $\Dlt\vphi$--squeezing and localization of
the wave function $\la\vphi|\psi\ra$: in view of the identification of
points $\vphi$ and $\vphi+2\pi$ the meanvalues $\la\vphi\ra$,
$\la\vphi^2\ra$ should be calculated by integration from $\vphi_0-\pi$ to
$\vphi_0+\pi$, where $\vphi_0$ is the center of the wave packet
(i.e. $\vphi_0$ is the most probable value of $\vphi$). In this way we
find that both standard deviations $\Dlt \vphi$ and $\Dlt J$ in K-R-P CS's
$|z\ra$ show very small oscillations around the value of $1/2$. So, the
family $\{|z\ra\}$ consists of almost minimum uncertainty states on the
circle.
\vspace{2mm}

{\bf\large 4.}
The {\it minimization states} (intelligent, or minimum-uncertainty states)
of the generalized UR (\ref{gSUR}) for $X_1$ and $X_2$ should
be eigenstates of a real or complex combination $\mu X_1 + \nu X_2$.
In the case of the
particle on a circle and $X_1=\hat{J}$ and $X_2=\hat{\vphi}$ the
$2\pi$--periodicity condition on the wave functions $\psi(\vphi+2\pi) =
\psi(\vphi)$ should be imposed (some authors admit exceptions \ci{D}). 
This restriction rules out all solutions of the eigenvalue equation 
$(\mu\hat J +\nu\hat\vphi)|\psi\ra = z|\psi\ra$, except for the eigenstates 
$\psi_m(\vphi)$ of $\hat J$, $\psi_m(\vphi) = (1/\sqrt{2\pi})\exp(im\vphi)$. 
For $\psi_m(\vphi)$ we have $\Dlt J = 0$, $\Dlt \vphi = \sqrt{\pi}$,
$G_{J\vphi} = 0$, so that the equality in UR (\ref{gSUR}) reads $0=0$.
None of the states $|z,\,a\ra$ and $|z\ra_s$ minimize the inequalities
(\ref{gSUR}), although the deviations in the case of CS's $|z\ra$ are very
small, as it is illustrated in figure 3 at $a=0$.

In order to define {\it squeezed states} on the circle let us recall that
for the particle on the real line these states are defined by means of one
of the two inequalities $(\Dlt x)^2 < |\la[x,p]\ra|/2 = 1/2$, or
$(\Dlt p)^2 < |\la[x,p]\ra|/2 = 1/2$. Since Im$G_{12}(\psi)$ is a
generalization of the mean commutator $(-i/2)\la[X_1,X_2]\ra$
one can define $X_1$-$X_2$ squeezed states more generally as states, for
which
\beq\lb{SS}  %eq. (17)
(_g\Dlt X_i)^2 < |{\rm Im}G_{12}(\psi)|,\quad i=1\,\, \mbox{or}\,\, 2.
\eeq
This is a generalization of the well known Eberly-Wodkiewicz criterion for
squeezed states. It however is a relative criterion, since the
"generalized mean commutator" $|{\rm Im}G_{12}(\psi)|$ may take, in
general, values from $0$ to $\infty$. Another and stronger criterion for
squeezed states is suggested by the observation that on the real line (and
for the one mode electromagnetic field) $1/2$ is the minimal value that two
variances $(\Dlt x)^2$ and $(\Dlt p)^2$ can take simultaneously. Therefore
we can define $X_1$-$X_2$ squeezed states more generally as states, for
which one of the following two inequalities holds,
\beq\lb{SS2}  %eq. (18)
(_g\Dlt X_i)^2 < \Dlt_0^2,\quad i=1\, \mbox{or}\, 2,
\eeq
where $\Dlt_0^2$ is the minimal value that the two generalized variances
can take simultaneously. For incompatible observables $\Dlt_0 > 0$.
It is plausible that $2\Dlt_0^2$ is the lower limit of the sum of two
variances,
\beq\lb{Dlt sum}
(\Dlt X_1)^2 + (\Dlt X_2)^2 \geq 2\Dlt_0^2.
\eeq
If the eigenstates of $X_1+iX_2$ (or $X_1-iX_2$) exist (canonical
observables, spin and quasi-spin components e.t.c.), then $\Dlt_0^2$ is
equal to the minimal value of $|{\rm Im}G_{J\vphi}(\psi)|$ within 
{\it these} eigenstates, and (\ref{Dlt sum}) is rigorously valid \ci{T2}. 
If eigenstates of $X_1\pm iX_2$  do not exist, the critical quantity
$\Dlt_0$ should be estimated by different methods. The case of $X_1= \hat
J$ and $X_2=\hat\vphi$ is such a special case, since $2\pi$--periodic
eigenfunctions of $\hat\vphi\pm i\hat J$ do not exist. Numerical
considerations suggest that in this case $\Dlt_0^2\approx 0.5$ (more
precisely $\approx 0.49999$), which is the minimal value that $(\Dlt
\vphi)^2$ and $(\Dlt J)^2$ take simultaneously in CS's $|z\ra$.

It turned out that both criterions (\ref{SS}) and (\ref{SS2}) can be
satisfied in many states from the families $\{|z,\,a\ra\}$ and
$\{|z\ra_s\}$.  Squeezing of $\Dlt\vphi$ in $|z,\,a\ra$ is
not very strong,  while in $|z\ra_s$ it can be arbitrarily strong.

Of course $|z\ra$ are exact Heisenberg intelligent states for the
Hermitian components $X$, $Y$ of $Z$.  However neither $\Dlt X$ nor $\Dlt
Y$ is in a satisfactory correspondence with the localization on a circle,
as one can easily check it on the example of cat states $|z;\pm\ra$.

In conclusion we note that the above described scheme can be extended to
represent correct generalized UR's for several observables and (several)
mixed states as well \ci{T}. 

\newpage

\begin{figure}
\centering
\includegraphics[width=100mm,height=45mm]{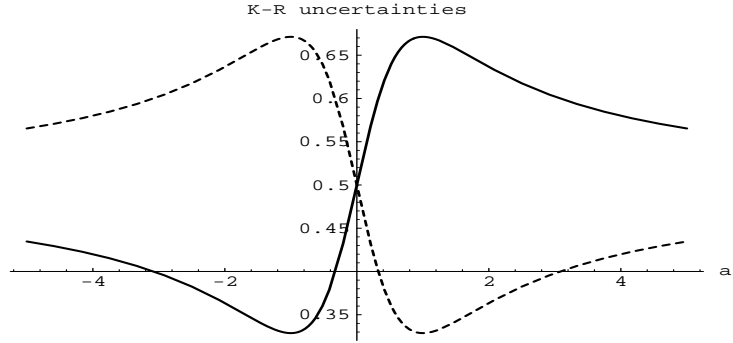}
\caption{K--R uncertainties
$\td\Dlt^2(\hat{\vphi})$ (solid line) and
$\td\Dlt^2(\hat{J})$ (dashed line) in cat states $|z\!=\!1,\,a\ra$ as
functions of $a$. $\td\Dlt^2(\hat{\vphi})$--squeezing is maximal around
$a=-1$.} 
\end{figure}

\begin{figure}
\centering
\includegraphics[width=100mm,height=45mm]{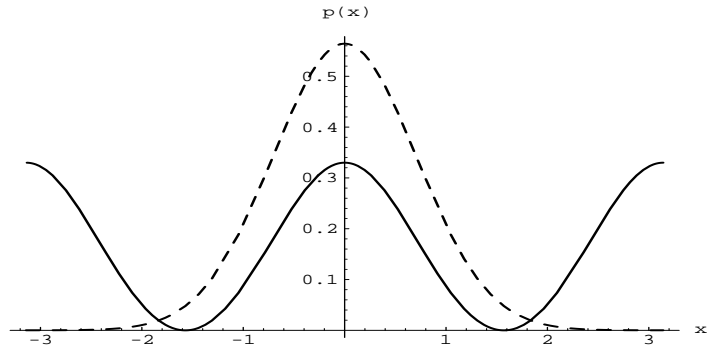}
\caption{The distributions $p(x)$ as functions of the angle $\vphi\equiv
x$ for $\td\Dlt^2(\hat{\vphi})$--squeezed state $|1;-\ra$ (solid line) and
non-squeezed CS $|z\!=\!1\ra$ (dashed line). CS $|1\ra$ is better
localized than $|1;-\ra$.} 
\end{figure}

\begin{figure}
\centering
\includegraphics[width=100mm,height=45mm]{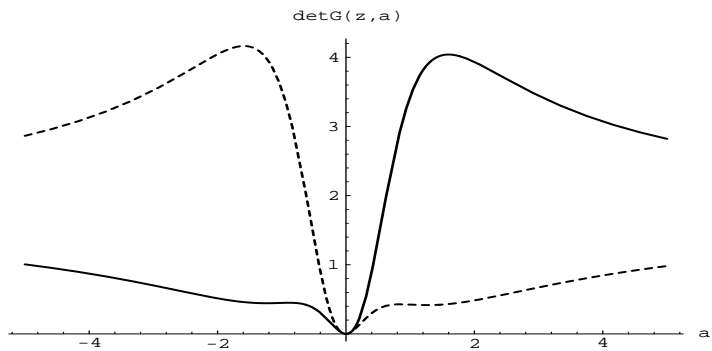}
\caption{Illustration of the generalized Schr\"odinger uncertainty
relation (10)  in the superpositions $|z,a\ra$:
$\det G = (\Dlt\vphi)^2(\Dlt J)^2 - |G_{J\vphi}|^2$ as a function of $a$
for $z=0.4$ (solid line) and $z=1$ (dashed line). }
\end{figure}

\end{document}